\documentclass[conference]{IEEEtran}
\usepackage[T1]{fontenc}
\usepackage[latin9]{inputenc}
\usepackage{bm}
\usepackage{amsmath}
\usepackage{amsthm}
\usepackage{amssymb}
\usepackage{graphicx}
\usepackage[unicode=true,
 bookmarks=true,bookmarksnumbered=true,bookmarksopen=true,bookmarksopenlevel=1,
 breaklinks=false,pdfborder={0 0 0},pdfborderstyle={},backref=false,colorlinks=true]
 {hyperref}
\hypersetup{pdftitle={Your Title},
 pdfauthor={Your Name},
 pdfpagelayout=OneColumn, pdfnewwindow=true, pdfstartview=XYZ, plainpages=false}

\makeatletter

\providecommand{\tabularnewline}{\\}

\theoremstyle{plain}
\newtheorem{thm}{\protect\theoremname}
\theoremstyle{definition}
\newtheorem{defn}[thm]{\protect\definitionname}
\theoremstyle{plain}
\newtheorem{lem}[thm]{\protect\lemmaname}
\theoremstyle{remark}
\newtheorem{rem}[thm]{\protect\remarkname}

\usepackage[caption=false,font=footnotesize]{subfig}
\usepackage{diagbox}
\usepackage{algorithm}
\usepackage{algorithmic}
\usepackage{color}

\IEEEoverridecommandlockouts
\usepackage[left=0.61in, right=0.61in, bottom=0.98in, top=0.75in]{geometry}

\@ifundefined{showcaptionsetup}{}{%
 \PassOptionsToPackage{caption=false}{subfig}}
\usepackage{subfig}
\makeatother

\providecommand{\definitionname}{Definition}
\providecommand{\lemmaname}{Lemma}
\providecommand{\remarkname}{Remark}
\providecommand{\theoremname}{Theorem}

\begin{document}
\title{Hybrid \nolinebreak Far- \nolinebreak and \nolinebreak Near-Field \nolinebreak Channel \nolinebreak Estimation \nolinebreak for THz \nolinebreak Ultra-Massive \nolinebreak MIMO \nolinebreak via \nolinebreak Fixed \nolinebreak Point \nolinebreak Networks}
\author{\author{\IEEEauthorblockN{Wentao~Yu, Yifei~Shen, Hengtao~He, Xianghao~Yu, Jun~Zhang, \textit{Fellow, IEEE}, and Khaled~B. Letaief, {\textit{Fellow, IEEE}}} 	\IEEEauthorblockA{ 		Dept. of ECE, The Hong Kong University of Science and Technology, Kowloon, Hong Kong\\ 		Email: \{wyuaq, yshenaw\}@connect.ust.hk, \{eehthe, eexyu, eejzhang, eekhaled\}@ust.hk}   \thanks{This work was supported by the Hong Kong Research Grants Council under Grant No. 16212120 and 15207220.} } }
\maketitle
\begin{abstract}
Terahertz ultra-massive multiple-input multiple-output (THz UM-MIMO)
is envisioned as one of the key enablers of 6G wireless systems. Due
to the joint effect of its array aperture and small wavelength, the
near-field region of THz UM-MIMO is greatly enlarged. The high-dimensional
channel of such systems thus consists of a stochastic mixture of far
and near fields, which renders channel estimation extremely challenging.
Previous works based on uni-field assumptions cannot capture the hybrid
far- and near-field features, thus suffering significant performance
loss. This motivates us to consider hybrid-field channel estimation.
We draw inspirations from fixed point theory to develop an efficient
deep learning based channel estimator with adaptive complexity and
linear convergence guarantee. Built upon classic orthogonal approximate
message passing, we transform each iteration into a contractive mapping,
comprising a closed-form linear estimator and a neural network based
non-linear estimator. A major algorithmic innovation involves applying
fixed point iteration to compute the channel estimate while modeling
neural networks with arbitrary depth and adapting to the hybrid-field
channel conditions. Simulation results verify our theoretical analysis
and show significant performance gains over state-of-the-art approaches
in the estimation accuracy and convergence rate. Source code is publicly available on \href{https://github.com/wyuaq/FPN-OAMP-THz-Channel-Estimation}{Github}. 
\end{abstract}

\IEEEpeerreviewmaketitle{}

\section{Introduction}

Terahertz (THz) band is the last piece of the radio frequency (RF)
spectrum puzzle for wireless systems \cite{2021Sarieddeen,2022Akyildiz}.
The massive bandwidth promises to provide ultra-high data rates and
seamlessly support new applications in 6G such as extended reality,
autonomous driving, and edge intelligence \cite{2022Letaief}. At
the same time, many problems, such as the limited coverage range,
are yet to be solved to fully unleash its potential. To combat the
coverage problem in an energy-efficient manner, ultra-massive multiple-input
multiple-output (UM-MIMO) with an array-of-subarray (AoSA) structure
has been proposed as a promising solution \cite{2021Sarieddeen,2022Ning}.
This structure groups the antenna array into multiple subarrays, each
powered by one RF chain, to perform highly directional hybrid beamforming
\cite{2016Yu,2020Zhang}. The fine-grained feature of beamforming
design necessitates accurate channel estimation with a low pilot overhead,
which, however, is extremely challenging due to limited RF chains. 

Conventional wireless systems operating at sub-6 GHz and millimeter
wave bands mostly considered the far-field region only, as the radius
of the near-field region, determined by the Rayleigh distance, is
much smaller compared with the coverage range. By contrast, for THz
UM-MIMO, the near-field region becomes critical due to the enlarged
Rayleigh distance, which, for example, is about $20$ m for an array
with $0.1$ m aperture at $300$ GHz. This will occupy a large portion
of the coverage of a typical THz system. Therefore, depending on the
distances between the RF source/scatterers and the array, far- and
near-field paths typically co-exist and together constitute the hybrid-field
channel. Considering such a unique feature, channel estimation algorithms
for THz UM-MIMO have to be compatible with both the far- and near-field
paths, and be robust against the variable channel \nolinebreak conditions. 

Unfortunately, so far there is no unified algorithm that can address
these challenges. To reduce pilot overhead, existing works mostly
adopted the uni-field assumption and exploited dedicated sparsity
patterns in either the far field \cite{2016Lee,2021Dovelos} or near
field \cite{2020Han} to design efficient compressed sensing algorithms.
A hybrid-field scenario was considered in \cite{2022Wei}, but the
authors assumed \textit{a priori} knowledge of whether each path is
from the far- or near-field region to decide what algorithm to apply,
which is far from practical. As an alternative, deep unfolding (DU)
methods can be adopted to learn the complex channel conditions by
augmenting classic iterative algorithms with learnable components
\cite{2018He,2019He,2020He,2022Ma,2022Shen}. Nevertheless, several
critical problems remain unsolved and hinder their application. Specifically,
although adapted from classic algorithms, the convergence of DU methods
is generally not guaranteed. In addition, DU methods are truncated
to a \textit{fixed} number of iterations. This contradicts the property
of classic iterative algorithms and can lead to an unstable performance
in the changeable channel conditions. 

\begin{figure*}[t]
\centering{}\subfloat[\label{fig:Array-of-subarray}]{
\centering{}\includegraphics[width=0.3\textwidth]{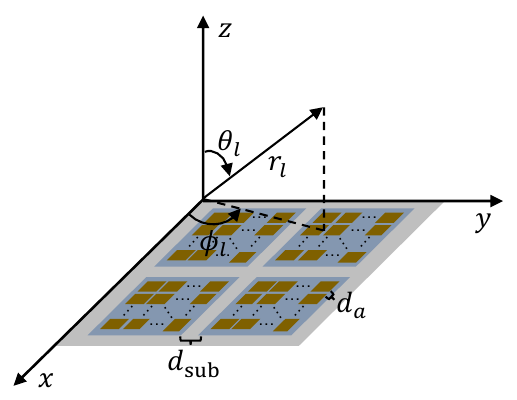}}\subfloat[\label{fig:Partially-connected}]{\centering{}\includegraphics[width=0.3\textwidth]{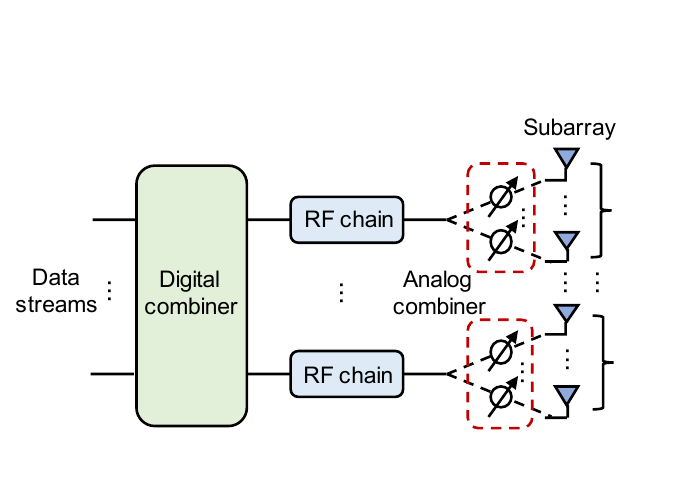}}\subfloat[\label{fig:Hybrid-field}]{\centering{}\includegraphics[width=0.3\textwidth]{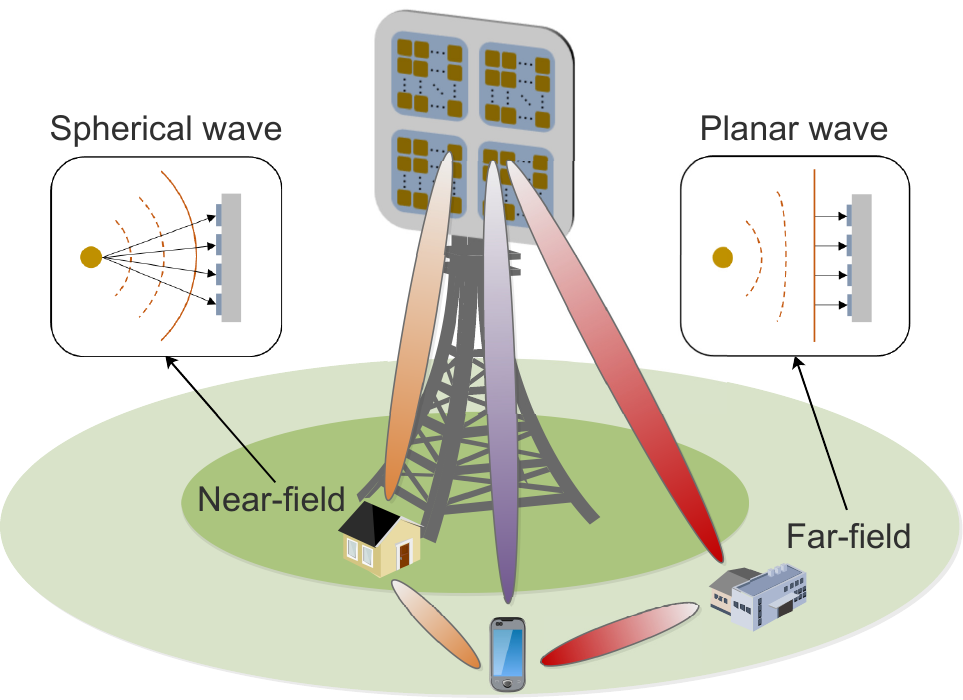}}\caption{System model. (a) Planar AoSA geometry of the THz UM-MIMO, in which
SAs are denoted by dark blue squares while AEs are denoted by dark
golden squares. (b) Partially-connected hybrid analog-digital beamforming
implemented by the AoSA, where the AEs in each SA share the same RF
chain through dedicated phase shifters. (c) A typical hybrid far-
and near-field propagation environment in THz UM-MIMO. \label{fig:System-model}}
\end{figure*}

To tackle these issues, we develop a deep learning based hybrid-field
channel estimator for THz UM-MIMO that enjoys convergence guarantee
and adaptive complexity. Specifically, inspired by fixed point theory
\cite{2019Bauschke}, we transform each iteration of orthogonal approximate
message passing (OAMP) \cite{2017Ma} into a contractive mapping,
by replacing the nonlinear estimator with a specially-trained convolutional
neural network (CNN). Thanks to the powerful modeling capacity of
CNNs, the patterns of the hybrid-field THz UM-MIMO channel can be
accurately identified and exploited. The estimated channel is computed
via a fixed point iteration of the contractive mapping. It is shown
that the proposed estimator enjoys provable linear convergence and
can model neural networks with the depth that adapts to the hybrid-field
channel conditions and an adjustable error tolerance. Simulation results
will verify our theoretical results and demonstrate that the proposed
method outperforms state-of-the-art approaches by a large margin. 

\textit{Notation:} Throughout this paper, $\mathbf{A}^{T}$, $\mathbf{A}^{H}$,
$\mathbf{A^{\dagger}}$, $\text{tr}(\mathbf{A})$, $\text{vec}(\mathbf{A})$,
and $(\mathbf{A})_{i,j}$ are respectively the transpose, Hermitian,
pseudoinverse, trace, vectorization, and the $(i,j)$-th element of
matrix $\mathbf{A}$. $\|\mathbf{a}\|_{p}$ and $(\mathbf{a})_{i}$
are the $\ell_{p}$-norm and the $i$-th element of vector $\mathbf{a}$,
respectively. $|a|$ is the absolute value of scalar $a$; $\mathbf{B}=\text{blkdiag}(\mathbf{A}_{1},\mathbf{A}_{2},\ldots,\mathbf{A}_{n})$
returns a block diagonal matrix by aligning $\mathbf{A}_{1},\mathbf{A}_{2},\ldots,\mathbf{A}_{n}$
along the diagonal. $\mathbf{I}$ and $\mathbf{0}$ are the identity
matrix and the all-zero vector of appropriate dimensions. $\mathbb{E}\{\cdot\}$
denotes expectation. $\circ$ denotes the composition of functions.
$\mathcal{U}(a_{1},a_{2})$ is a continuous uniform distribution over
the interval of $a_{1}$ and $a_{2}$. $\mathcal{CN}(\boldsymbol{\mu},\mathbf{R})$
is a complex normal distribution with mean $\boldsymbol{\mu}$ and
covariance $\mathbf{R}$. 

\section{System Model and Problem Formulation}

We consider the uplink channel estimation for THz UM-MIMO systems.
The base station (BS) is equipped with a planar AoSA with $\sqrt{S}\times\sqrt{S}$
subarrays (SAs), while each SA is a uniform planar array consisting
of $\sqrt{\bar{S}}\times\sqrt{\bar{S}}$ antenna elements (AEs), as
illustrated in Fig. \ref{fig:System-model}(a). To improve energy
efficiency, the AoSA adopts partially-connected hybrid analog-digital
beamforming \cite{2020Zhang}, as shown in Fig. \ref{fig:System-model}(b).
Within each SA, the AEs share the same RF chain through dedicated
phase shifters. A total of $S$ RF chains are utilized to receive
data streams from multiple single-antenna user equipments (UEs). 

We define the index $s$ of the SA at the $m$-th row and $n$-th
column of the AoSA by $s=(m-1)\sqrt{S}+n$, where $1\leq m,n\leq\sqrt{S}$
and $1\leq s\leq S$. Similarly, the index $\bar{s}$ of the AE at
the $\bar{m}$-th row and $\bar{n}$-th column of a certain SA is
defined by $\bar{s}=(\bar{m}-1)\sqrt{\bar{S}}+\bar{n}$, where $1\leq\bar{m},\bar{n}\leq\sqrt{\bar{S}}$
and $1\leq\bar{s}\leq\bar{S}$. The distances between adjacent SAs
and adjacent AEs are denoted by $d_{\text{sub}}$ and $d_{a}$, respectively.
As shown in Fig. \ref{fig:System-model}(a), we construct a Cartesian
coordinate system with the origin being the first AE in the first
SA. Assuming that the AoSA lies in the $x$-$y$ plane, then the coordinate
of the $\bar{s}$-th AE in the $s$-th SA is given by

\noindent\begin{minipage}[c][0.9\totalheight]{1\columnwidth}%
\begin{equation}
\mathbf{p}_{s,\bar{s}}=\left(\begin{array}{c}
(m-1)[(\sqrt{\bar{S}}-1)d_{a}+d_{\text{sub}}]+(\bar{m}-1)d_{a}\\
(n-1)[(\sqrt{\bar{S}}-1)d_{a}+d_{\text{sub}}]+(\bar{n}-1)d_{a}\\
0
\end{array}\right).
\end{equation}
\end{minipage}

\subsection{Hybrid-Field THz UM-MIMO Channel Model}

Here we introduce the hybrid far- and near-field propagation environment
along with channel model. The boundary of the far- and near-field
regions is determined by the Rayleigh distance, i.e., $D_{\text{Rayleigh}}=\frac{2D^{2}}{\lambda_{c}}$,
where $D$ is the array aperture, and $\lambda_{c}$ is the carrier
wavelength. Due to the enlarged near-field region in THz UM-MIMO,
the channel can consist of both the far- and near-field paths, as
shown in Fig. \ref{fig:System-model}(c). The number of far- and near-field
paths may vary, which renders the channel condition changeable. Additionally,
since the wavefront is approximately planar in the far field and spherical
in the near field, the array responses should be modeled separately. 

Due to limited scattering, the spatial channel $\mathbf{\tilde{h}}\in\mathbb{C}^{S\bar{S}\times1}$
between the BS and a specific UE can be characterized by the superposition
of one LoS path and $L-1$ NLoS paths \cite{2021Dovelos}, i.e.,

\noindent\begin{minipage}[c][1.5\totalheight]{1\columnwidth}%
\begin{equation}
\mathbf{\tilde{h}}=\gamma\sum_{l=1}^{L}\alpha_{l}\mathbf{a}\left(\phi_{l},\theta_{l},r_{l}\right)e^{-j2\pi f_{c}\tau_{l}}\label{eq:channel}
\end{equation}
\end{minipage}

\noindent where $\gamma$ is a normalization factor such that $\|\mathbf{\tilde{h}}\|_{2}^{2}=S\bar{S}$,
$f_{c}$ is the carrier frequency. Also, $\alpha_{l}$, $\phi_{l}$,
$\theta_{l}$, $r_{l}$, $\mathbf{a}\left(\phi_{l},\theta_{l},r_{l}\right)$,
and $\tau_{l}$ are respectively the path loss, azimuth angle of arrival
(AoA), elevation AoA, distance between the array and the RF source/scatterer,
array response vector, and time delay of the $l$-th path. In particular,
$\phi_{l}$, $\theta_{l}$, $r_{l}$ are measured with respect to
the origin of the coordinate system, as shown in Fig. \ref{fig:System-model}(a). 

\subsubsection{Path Loss}

The path loss $\alpha_{l}$ accounts for both the spread loss and
the molecular absorption loss. Assuming that $l=1$ denotes the LoS
path and $l>1$ denote NLoS paths, then

\noindent\begin{minipage}[c][1.5\totalheight]{1\columnwidth}%
\begin{equation}
\alpha_{l}=|\Gamma_{l}|\left(\frac{c}{4\pi f_{c}r_{1}}\right)e^{-\frac{1}{2}k_{\text{abs}}r_{1}},
\end{equation}
\end{minipage}

\noindent where $\Gamma_{l}$ is the reflection coefficient, $r_{1}$
is the LoS path length, and $k_{\text{abs}}$ is the molecular absorption
coefficient \cite{2021Dovelos}. For the LoS path, $\Gamma_{l}=1$.
For the NLoS paths, $\Gamma_{l}$ is given by 

\noindent\begin{minipage}[c][1.5\totalheight]{1\columnwidth}%
\begin{equation}
\Gamma_{l}=\frac{\cos\varphi_{\text{in},l}-n_{t}\cos\varphi_{\text{ref},l}}{\cos\varphi_{\text{in},l}+n_{t}\cos\varphi_{\text{ref},l}}e^{-\left(\frac{8\pi^{2}f_{c}^{2}\sigma_{\text{rough}}^{2}\text{cos}^{2}\varphi_{\text{in},l}}{c^{2}}\right)},
\end{equation}
\end{minipage}

\noindent where $\varphi_{\text{in},l}$ is the angle of incidence
of the $l$-th path, $\varphi_{\text{ref},l}=\arcsin(n_{t}^{-1}\sin\varphi_{\text{in},l})$
is the angle of refraction. Also, $n_{t}$ and $\sigma_{\text{rough}}$
are respectively the refractive index and the roughness coefficient
of the reflecting material \cite{2021Dovelos}. 

\subsubsection{Array Response Vector}

The array response vector $\mathbf{a}^{\text{}}(\phi_{l},\theta_{l},r_{l})\in\mathbb{C}^{S\bar{S}\times1}$
differs in the far- and near-field regions, which are determined by
the distance $r_{l}$, and is given by

\noindent\begin{minipage}[c][1.5\totalheight]{1\columnwidth}%
\begin{equation}
\mathbf{a}(\phi_{l},\theta_{l},r_{l})=\begin{cases}
\mathbf{a}^{\text{far}}(\phi_{l},\theta_{l}), & \text{if }r_{l}>D_{\text{Rayleigh}},\\
\mathbf{a}^{\text{near}}(\phi_{l},\theta_{l},r_{l}) & \text{otherwise}.
\end{cases}
\end{equation}
\end{minipage}

\noindent For notational brevity, we first construct the array response
matrix. Due to the planar wavefront, each element of the far-field
array response matrix $\mathbf{A}^{\text{far}}(\phi_{l},\theta_{l})$
is

\noindent\begin{minipage}[c][1.5\totalheight]{1\columnwidth}%
\begin{equation}
(\mathbf{A}^{\text{far}}(\phi_{l},\theta_{l}))_{s,\bar{s}}=e^{-j2\pi\frac{f_{c}}{c}\mathbf{p}_{s,\bar{s}}^{T}\mathbf{t}_{l}},
\end{equation}
\end{minipage}

\noindent where $c$ is the speed of light, and $\mathbf{t}_{l}$
is the unit-length vector in the AoA direction of the $l$-th path,
given by $\mathbf{t}_{l}=(\sin\theta_{l}\cos\phi_{l},\sin\theta_{l}\sin\phi_{l},\cos\theta_{l})^{T}$.
The corresponding far-field array response vector is given by $\mathbf{a}^{\text{far}}(\phi_{l},\theta_{l})=\text{vec}(\mathbf{A}^{\text{far}}(\phi_{l},\theta_{l}))$.
Due to the spherical wavefront, each element of the near-field array
response matrix depends on the exact distance between the AE and the
RF source/scatterer, i.e., 

\noindent\begin{minipage}[c][1.5\totalheight]{1\columnwidth}%
\begin{equation}
(\mathbf{A}^{\text{near}}(\phi_{l},\theta_{l},r_{l}))_{s,\bar{s}}=e^{-j2\pi\frac{f_{c}}{c}\|\mathbf{p}_{s,\bar{s}}-r_{l}\mathbf{t}_{l}\|_{2}}.
\end{equation}
\end{minipage}

\noindent The near-field array response vector can be similarly obtained
by vectorization, i.e., $\mathbf{a}^{\text{near}}(\phi_{l},\theta_{l},r_{l})=\text{vec}(\mathbf{A}^{\text{near}}(\phi_{l},\theta_{l},r_{l}))$. 

\subsection{Problem Formulation}

In uplink channel estimation, the UEs transmit known pilot signals
to the BS for $Q$ time slots. We assume that orthogonal pilots are
adopted and consider an arbitrary UE without loss of generality. For
the ease of algorithm design and comparison, the spatial channel $\mathbf{\tilde{h}}$
is transformed to its angular domain representation $\mathbf{\bar{h}}$
in an SA-by-SA manner by using $\mathbf{\bar{h}}=\mathbf{F}^{H}\mathbf{\tilde{h}}$,
where $\mathbf{F}^{H}=\text{blkdiag}(\mathbf{U}_{1},\mathbf{U}_{2},\ldots,\mathbf{U}_{S})$
is a unitary matrix with each $\mathbf{U}_{s}$ being an $\bar{S}\times\bar{S}$
matrix constructed by the Kronecker product of two normalized discrete
Fourier transform matrices of size $\sqrt{\bar{S}}\times\sqrt{\bar{S}}$.
The received pilot signal $\mathbf{y}_{q}\in\mathbb{C}^{S\times1}$
in the $q$-th time slot is given by

\noindent\begin{minipage}[c][1.8\totalheight]{1\columnwidth}%
\begin{equation}
\mathbf{y}_{q}=\mathbf{W}_{\text{BB},q}^{H}\mathbf{W}_{\text{RF},q}^{H}\mathbf{F}\mathbf{\bar{h}}s_{q}+\mathbf{W}_{\text{BB},q}^{H}\mathbf{W}_{\text{RF},q}^{H}\mathbf{n}_{q},
\end{equation}
\end{minipage}

\noindent where $\mathbf{W}_{\text{BB},q}\in\mathbb{C}^{S\times S}$
is the digital combining matrix, $\mathbf{W}_{\text{RF},q}\mathbf{=\text{blkdiag\ensuremath{\left(\mathbf{w}_{1,q},\mathbf{w}_{2,q},\ldots\mathbf{w}_{S,q}\right)}}\in\mathbb{C}}^{S\bar{S}\times S}$
is the analog combining matrix where the elements of each component
vector $\mathbf{w}_{i,q}\in\mathbb{C}^{\bar{S}\times1}$ satisfy the
constant-modulus constraint, $s_{q}$ is the known pilot signal that
is set as 1 for convenience, and $\mathbf{\mathbf{n}}_{q}\sim\mathcal{CN}(\mathbf{0},\sigma_{n}^{2}\mathbf{I})$
is the noise. The average received signal-to-noise-ratio (SNR) is
$\frac{1}{\sigma_{n}^{2}}$. Since the combining matrices cannot be
optimally tuned without knowledge of the channel, we consider an arbitrary
scenario where $\mathbf{W}_{\text{BB},q}$ is set as the identity
matrix $\mathbf{I}$ and the analog phase shifts in $\mathbf{W}_{\text{RF},q}$
are randomly chosen from one-bit quantized angles, i.e., $(\mathbf{w}_{i,q})_{j}\in\frac{1}{\sqrt{S\bar{S}}}\{\pm1\}$,
to reduce energy consumption \cite{2018He}. The received signal $\mathbf{\bar{y}}=[\mathbf{y}_{1}^{T},\mathbf{y}_{2}^{T},\ldots,\mathbf{y}_{Q}^{T}]^{T}\in\mathbb{C}^{SQ\times1}$
after $Q$ time slots of pilot transmission is given by

\noindent\begin{minipage}[c][1.8\totalheight]{1\columnwidth}%
\begin{equation}
\mathbf{\bar{y}}=\text{\ensuremath{\mathbf{\bar{M}}}}\mathbf{\bar{h}}+\mathbf{\bar{n}},\label{eq:complex_inverse}
\end{equation}
\end{minipage}

\noindent where $\bar{\mathbf{M}}=[(\mathbf{W}_{\text{RF},1}^{H}\mathbf{F})^{T},\ldots,(\mathbf{W}_{\text{RF},Q}^{H}\mathbf{F})^{T}]^{T}\in\mathbb{C}^{SQ\times S\bar{S}}$,
and $\mathbf{\bar{n}}=[(\mathbf{W}_{\text{RF},1}^{H}\mathbf{n}_{1})^{T},\ldots,(\mathbf{W}_{\text{RF},Q}^{H}\mathbf{n}_{Q})^{T}]^{T}\in\mathbb{C}^{SQ\times1}$. 

To transform \eqref{eq:complex_inverse} into its equivalent real-valued
form, we let $\mathbf{y}=[\text{\ensuremath{\Re}}(\mathbf{\bar{y}})^{T},\text{\ensuremath{\Im}}(\mathbf{\bar{y}})^{T}]^{T}\in\mathbb{R}^{2SQ\times1}$,
$\mathbf{h}=[\text{\ensuremath{\Re}}(\mathbf{\bar{h}})^{T},\text{\ensuremath{\Im}}(\mathbf{\bar{h}})^{T}]^{T}\in\mathbb{R}^{2S\bar{S}\times1}$.
$\mathbf{n}=[\text{\ensuremath{\Re}}(\mathbf{\bar{n}})^{T},\text{\ensuremath{\Im}}(\mathbf{\bar{n}})^{T}]^{T}\in\mathbb{R}^{2SQ\times1}$,
and 

\noindent\begin{minipage}[c][1.5\totalheight]{1\columnwidth}%
\begin{equation}
\mathbf{M}=\left(\begin{array}{cc}
\text{\ensuremath{\Re}}(\mathbf{\bar{M}}) & -\text{\ensuremath{\Im}}(\mathbf{\bar{M}})\\
\text{\ensuremath{\Im}}(\mathbf{\bar{M}}) & \text{\ensuremath{\Re}}(\mathbf{\bar{M}})
\end{array}\right)\in\mathbb{R}^{2SQ\times2S\bar{S}}.
\end{equation}
\end{minipage}

\noindent Then, the equivalent real-valued form is given by

\noindent\begin{minipage}[b][1.5\totalheight][c]{1\columnwidth}%
\begin{equation}
\mathbf{\mathbf{y}=Mh+n}.\label{eq:real-inverse}
\end{equation}
\end{minipage}

Based on \eqref{eq:real-inverse}, channel estimation can be formulated
as a linear inverse problem whose goal is to compute a good estimate
of $\mathbf{h}$ given the knowledge of $\mathbf{y}$ and $\mathbf{M}$.
However, due to the practical requirements of low pilot overhead and
limited RF chains, it is often the case that $SQ\ll S\bar{S}$, which
makes the problem significantly ill-posed. Existing works rely heavily
on the channel sparsity to design compressed sensing algorithms for
channel estimation. However, the sparsifying transformations for the
far- and near-field paths are not compatible with each other \cite{2022Wei}.
DU methods could be adopted but they generally lack theoretical guarantees.
Additionally, the \textit{fixed} number of layers limits their ability
to adapt to the changeable channel conditions and can cause unstable
performance. These drawbacks motivate us to design an efficient deep
learning based hybrid-field channel estimator with provable convergence
guarantee and adaptive complexity. 

\section{Fixed Point Networks for Hybrid-Field THz UM-MIMO Channel Estimation}

\begin{figure*}[tbh]
\centering{}\subfloat[]{\centering{}\includegraphics[height=0.09\textwidth]{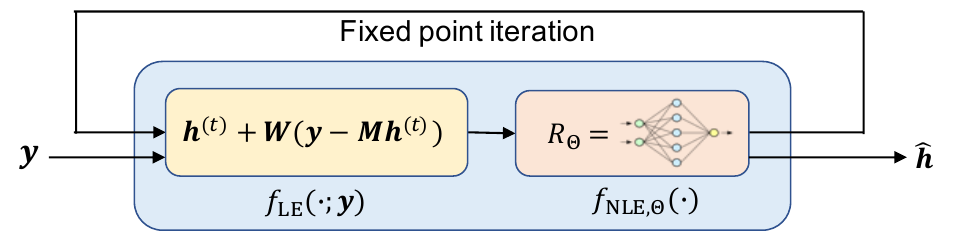}}\subfloat[]{\centering{}\includegraphics[height=0.112\textwidth]{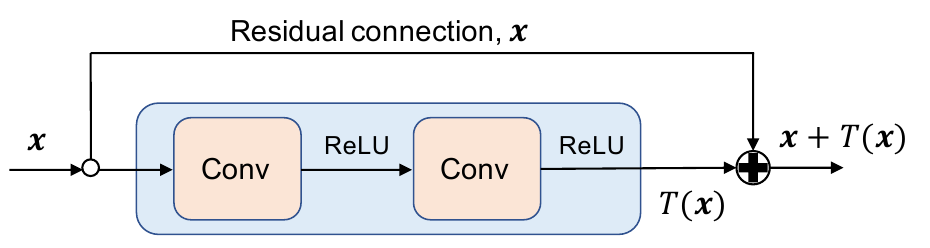}}\caption{(a) The schematic diagram of the proposed FPN-OAMP. (b) The structure
of one residual block. Let the input be $\mathbf{x}$, then the output
is $\mathbf{x+}T(\mathbf{x})$. \label{fig:FPN-diagram}}
\end{figure*}

\subsection{Fixed Point Modeling of Neural Networks}

Most iterative algorithms for solving linear inverse problems can
be represented in the following general form, i.e.,

\noindent\begin{minipage}[b][1.5\totalheight][c]{1\columnwidth}%
\begin{equation}
\mathbf{h}^{(t+1)}=f_{\Theta}(\mathbf{h}^{(t)};\mathbf{y}),\label{eq:general-update}
\end{equation}
\end{minipage}

\noindent where $\mathbf{h}^{(t)}$ denotes the intermediate estimation
at the $t$-th iteration, and $f_{\Theta}$ denotes a mapping parameterized
by $\Theta$. Well-known examples of this general form include proximal
algorithms, OAMP, and also weight-tied neural networks. The limit
as $t\rightarrow\infty$, i.e., $\mathbf{h}^{*}$, supposed that it
exists, is a solution of the fixed point equation

\noindent\begin{minipage}[c][1.8\totalheight]{1\columnwidth}%
\begin{equation}
\mathbf{h}^{*}=f_{\Theta}(\mathbf{h}^{*};\mathbf{y}),\label{eq:fixed-point-eq}
\end{equation}
\end{minipage}

\noindent which models the behavior of an algorithm at its convergence.
Particularly, if $f_{\Theta}(\cdot;\mathbf{y})$ corresponds to one
layer of a weight-tied neural network, then the fixed point $\mathbf{h}^{*}$
is the output of the network after an \textit{infinite} number of
layers \cite{2019Bai,2022Fung}. 

Our basic idea is to design a neural network assisted mapping $f_{\Theta}(\cdot;\mathbf{y})$
such that its fixed point is a good estimate of the hybrid-field channel
given the pilot measurement $\mathbf{y}$. We refer to such a general
framework as the \textit{fixed point network} (FPN). Specifically,
$f_{\Theta}(\cdot;\mathbf{y})$ can be constructed by adding learnable
components to various different algorithms that belong to \eqref{eq:general-update}.
The remaining question is how we can ensure the existence of the fixed
point and find it efficiently. Before further discussion, we first
define two key concepts. 
\begin{defn}[Lipschitz continuity]
A mapping $f_{\Theta}(\cdot;\mathbf{y})$ is Lipschitz continuous
if there exists a constant $L$ such that 

\noindent\begin{minipage}[b][1.5\totalheight][c]{1\columnwidth}%
\[
\|f_{\Theta}(\mathbf{h}_{1};\mathbf{y})-f_{\Theta}(\mathbf{h}_{2};\mathbf{y})\|\leq L\|\mathbf{h}_{1}-\mathbf{h}_{2}\|
\]
\end{minipage}

\noindent holds for any $\mathbf{h}_{1},\mathbf{h}_{2}\in\text{dom}(f_{\Theta}(\cdot;\mathbf{y}))$. 
\end{defn}
\begin{defn}[Contractive]
A mapping $f_{\Theta}(\cdot;\mathbf{y})$ is contractive if it is
Lipschitz continuous with constant $0\leq L<1$. 
\end{defn}
The existence of the fixed point and an efficient way to find it can
both be ensured by fixed point theory. As long as $f_{\Theta}(\cdot;\mathbf{y})$
is a contractive mapping (no matter what detailed operations it contains),
a simple repeated application of $f_{\Theta}(\cdot;\mathbf{y})$ will
make $\mathbf{h}^{(t)}$ converge linearly to the unique \nolinebreak fixed \nolinebreak point
$\mathbf{h}^{*}$. 
\begin{thm}[{Banach-Picard \cite[Theorem 1.50]{2019Bauschke}\label{Banach-Picard}}]
For any initial value $\mathbf{h}^{(0)}$, if the sequence $\{\mathbf{h}^{(t)}\}$
is generated via the relation $\mathbf{h}^{(t+1)}=f_{\Theta}(\mathbf{h}^{(t)};\mathbf{y})$
and $f_{\Theta}(\cdot;\mathbf{y})$ is a contractive mapping with
Lipschitz constant $0\leq L<1$, then $\{\mathbf{h}^{(t)}\}$ converges
to the unique fixed point $\mathbf{h}^{*}$ of $f_{\Theta}(\cdot;\mathbf{y})$
with a linear convergence rate $L$. The gap between $\mathbf{h}^{(t)}$
and $\mathbf{h}^{*}$ decreases geometrically as $\|\mathbf{h}^{(t+1)}-\mathbf{h}^{*}\|_{2}\leq L\|\mathbf{h}^{(t)}-\mathbf{\mathbf{h}}^{*}\|_{2}$. 
\end{thm}
This theorem reveals several unique advantages of the FPNs that are
not available in prevailing DU methods. First, it provides a simple
and unified framework to establish convergence guarantee. The only
requirement, i.e., $f_{\Theta}(\cdot;\mathbf{y})$ is contractive,
can be satisfied by controlling \nolinebreak the Lipschitz constant
of the neural network during training \cite{2021Gouk}. Second, the
complexity of FPNs is adaptive and can be adjusted at the testing
time. Since the fixed point iteration converges linearly to the unique
fixed point $\mathbf{h}^{*}$, one can run it to an arbitrary depth
depending on the desired accuracy and the hybrid-field channel condition
(reflected in $\mathbf{y}$). This will offer a flexible tradeoff
between complexity and performance, as well as an excellent ability
to adapt to the changeable channel conditions. 

\subsection{FPN-OAMP}

As mentioned before, there are a lot of possible choices for the mapping
$f_{\Theta}$. To incorporate wireless domain knowledge, we design
it based on the algorithmic structure of a powerful compressed sensing
algorithm, i.e., the OAMP. The proposed FPN-based variant of it, called
the FPN-OAMP, consists of a closed-form linear estimator (LE) and
a CNN-based nonlinear estimator (NLE). The mapping $f_{\Theta}(\cdot;\mathbf{y})$
is a composition of them, i.e., $f_{\Theta}(\cdot;\mathbf{y})=(f_{\text{NLE},\Theta}\circ f_{\text{LE}})(\cdot;\mathbf{y})$.
The process of FPN-OAMP is summarized in Fig. \ref{fig:FPN-diagram}(a)
and \textbf{Algorithm 1}. 

\subsubsection{Linear Estimator}

The LE of FPN-OAMP is similar to that of OAMP. It is given by

\noindent\begin{minipage}[c][1.5\totalheight]{1\columnwidth}%
\begin{equation}
f_{\text{LE}}(\mathbf{h}^{(t)};\mathbf{y})=\mathbf{h}^{(t)}+\mathbf{W}(\mathbf{y}-\mathbf{Mh}^{(t)}).
\end{equation}
\end{minipage}

\noindent The LE matrix $\mathbf{W}\in\mathbb{R}^{2S\bar{S}\times2SQ}$
is constructed by\footnote{The measurement matrix $\mathbf{M}$ is fixed since the combining
matrices cannot be optimally tuned without knowledge of the channel
\cite{2018He}. As a result, the LE matrix $\mathbf{W}$ is fixed,
and thus the computation of pseudoinverse is avoided. }

\noindent\begin{minipage}[c][1.5\totalheight]{1\columnwidth}%
\begin{equation}
\mathbf{W}=\eta\text{\ensuremath{\mathbf{M}^{\dagger}}}=\frac{2S\bar{S}}{\text{tr}(\mathbf{M}^{\dagger}\mathbf{M})}\mathbf{M}^{\dagger},
\end{equation}
\end{minipage}

\noindent where $\eta$ is the step size that ensures $\text{tr}(\mathbf{I}-\mathbf{WM})=0$,
such that the LE is de-correlated \cite{2017Ma}. 

\subsubsection{Nonlinear Estimator}

The NLE is given by

\noindent\begin{minipage}[c][1.5\totalheight]{1\columnwidth}%
\begin{equation}
f_{\text{NLE,\ensuremath{\Theta}}}(f_{\text{LE}}(\mathbf{h}^{(t)};\mathbf{y}))=R_{\Theta}(f_{\text{LE}}(\mathbf{h}^{(t)};\mathbf{y})),
\end{equation}
\end{minipage}

\noindent where $R_{\Theta}(\cdot)$ is a ResNet-like structure consisting
of three residual blocks (RBs), as shown in Fig. \ref{fig:FPN-diagram}(b).
Before the RBs, $f_{\text{LE}}(\mathbf{h}^{(t)};\mathbf{y})$ is first
reshaped into a tensor of $S$ feature maps of size $\sqrt{\bar{S}}\times\sqrt{\bar{S}}$,
each corresponding to an SA, and then passes through a convolution
(Conv) layer to lift them to 64 feature maps. Each RB constitutes
two Conv layers with $3\times3$ kernels and a fixed number of 64
feature maps, which are respectively followed by a ReLU activation.
We further follow the RBs by two $1\times1$ Conv layers, where the
first one adopts leaky ReLU activation, before reshaping back to the
vector form $\mathbf{h}^{(t+1)}$. 

\begin{algorithm} 
\caption{FPN-OAMP for hybrid-field channel estimation}
\begin{algorithmic}[1]
\STATE {\bf Input:} Measurement matrix $\mathbf{M}$, received pilot signals $\mathbf{y}$, weights of the nonlinear estimator $\Theta$, error tolerance $\epsilon$ \\
\STATE {\bf Output:} Estimated hybrid-field THz channel $\hat{\mathbf{h}}$ \\
\STATE {\bf Initialize:} $\mathbf{h}^{(0)} \gets \mathbf{0}, t \gets 0$ \\
\STATE \textcolor{black}{Fixed point iteration of $(f_{\text{NLE},\Theta} \circ f_{\text{LE}})(\cdot;\mathbf{y})$}:
\STATE {\bf while} $\|\mathbf{h}^{(t)}-(f_{\text{NLE},\Theta} \circ f_{\text{LE}})(\mathbf{h}^{(t)};\mathbf{y})\|_2 > \epsilon$ {\bf do}
\STATE\hspace{\algorithmicindent} $\mathbf{h}^{(t+1)} \gets (f_{\text{NLE},\Theta} \circ f_{\text{LE}})(\mathbf{h}^{(t)};\mathbf{y})$
\STATE\hspace{\algorithmicindent} $t \gets t+1$
\STATE $\hat{\mathbf{h}} \gets \mathbf{h}^{(t)}$
\STATE {\bf return} $\hat{\mathbf{h}}$  
\end{algorithmic}
\end{algorithm}

\subsection{Linear Convergence of the FPN-OAMP\label{subsec:Linear-Convergence}}

To prove the linear convergence of the FPN-OAMP, on the basis of \textbf{Theorem
\ref{Banach-Picard}}, we need to show that $(f_{\text{NLE},\Theta}\circ f_{\text{LE}})(\cdot;\mathbf{y})$
is contractive, which is proved as follows. 
\begin{lem}[\cite{2021Gouk}]
The composition of an $L_{1}$-Lipschitz and an $L_{2}$-Lipschitz
mapping is $L_{1}L_{2}$-Lipschitz. 
\end{lem}
\begin{thm}[]
Each layer of the FPN-OAMP $(f_{\text{NLE},\Theta}\circ f_{\text{LE}})(\cdot;\mathbf{y})$
is a contractive mapping if $R_{\Theta}(\cdot)$ is contractive. 
\end{thm}
\begin{IEEEproof}
We begin by showing that the Lipschitz constant of $f_{\text{LE}}(\cdot;\mathbf{y})$
is 1. Because the non-zero eigenvalues of $\mathbf{M}^{\dagger}\mathbf{M}$
and $\mathbf{MM}^{\dagger}=\mathbf{I}$ are the same, the eigenvalues
of $\mathbf{M}^{\dagger}\mathbf{M}$ equal either 0 or 1. Since $f_{\text{LE}}(\cdot;\mathbf{y})$
is an affine mapping, its Lipschitz constant is the spectral norm
of $\mathbf{I}-\eta\mathbf{M}^{\dagger}\mathbf{M}$, i.e., the largest
singular value of the matrix, given by

\noindent\begin{minipage}[c][1.5\totalheight]{1\columnwidth}%
\begin{equation}
\text{Lip}(f_{\text{LE}}(\cdot;\mathbf{y}))=\max_{i}(1-\eta\lambda_{i}(\mathbf{M}^{\dagger}\mathbf{M}))=1,
\end{equation}
\end{minipage}

\noindent where $\lambda_{i}(\cdot)$ denotes the $i$-th eigenvalue
of a matrix. Therefore, according to \textbf{Lemma 4}, the composition
$(f_{\text{NLE},\Theta}\circ f_{\text{LE}})(\cdot;\mathbf{y})$ is
a contractive mapping if $R_{\Theta}(\cdot)$ is contractive. 
\end{IEEEproof}
\begin{rem}
We provide details on training a contractive $R_{\Theta}(\cdot)$
afterwards. Since $(f_{\text{NLE},\Theta}\circ f_{\text{LE}})(\cdot;\mathbf{y})$
is contractive regardless of $\mathbf{y}$, the linear convergence
rate of FPN-OAMP will hold for different channel conditions and SNR
levels. The error tolerance $\epsilon$ in \textbf{Algorithm 1} explicitly
controls the accuracy of the approximate fixed point $\mathbf{\hat{h}}$,
since the gap between $\mathbf{h}^{(t)}$ and $\mathbf{h}^{*}$ decreases
geometrically. Adjusting $\epsilon$ can provide a flexible tradeoff
between complexity and performance. 
\end{rem}

\subsection{Training of the FPN-OAMP\label{subsec:Training-the-FPN-OAMP}}

During training, we first run the fixed point iteration to find the
approximate fixed point $\hat{\mathbf{h}}$ given an error tolerance
$\epsilon$. The loss function is chosen as the normalized mean squared
error (NMSE) of $\mathbf{\hat{h}}$ and the ground truth channel $\mathbf{h}$,
i.e.,

\noindent\begin{minipage}[c][1.5\totalheight]{1\columnwidth}%
\begin{equation}
\ell(\mathbf{\hat{h}},\mathbf{h})\triangleq\mathbb{E}\{\|\mathbf{h}-\mathbf{\hat{h}}\|_{2}^{2}/\|\mathbf{h}\|_{2}^{2}\}.
\end{equation}
\end{minipage}

\noindent We adopt the Jacobian-free backpropagation in \cite{2022Fung}
to train the FPN-OAMP, which only imposes a \textit{constant} memory
overhead regardless of the number of fixed point iterations. To enforce
the contractive property of $R_{\Theta}(\cdot)$, we check its Lipschitz
constant after each weight update using the current batch of training
data. The Lipschitz constant is approximated by

\noindent\begin{minipage}[c][1.5\totalheight]{1\columnwidth}%
\begin{equation}
L=\frac{{\scriptstyle {\displaystyle {\scriptstyle \sum_{i=1}^{B}\|R_{\Theta}(\text{\ensuremath{\mathbf{\hat{h}}_{i}}}\mathbf{+\bm{\delta}}_{i};\mathbf{y})-R_{\Theta}(\mathbf{\text{\ensuremath{\mathbf{\hat{h}}_{i}}}};\mathbf{y})\|_{2}}}}}{{\scriptstyle \sum_{i=1}^{B}\|\mathbf{\bm{\delta}}_{i}\|_{2}}},\label{eq:Lipschitz}
\end{equation}
\end{minipage}

\noindent where $B$ is the batch size, $\text{\ensuremath{\mathbf{\hat{h}}_{i}}}$
corresponds to the $i$-th training sample, and $\bm{\delta}_{i}$
is a small random perturbation \cite{2021Heaton}. According to \textbf{Lemma
4}, since the LE and the (leaky) ReLU activations in the NLE are all
1-Lipschitz \cite{2021Gouk}, the Lipschitz constant of $(f_{\text{NLE},\Theta}\circ f_{\text{LE}})(\cdot;\mathbf{y})$
is only determined by the Conv layers. Additionally, since each Conv
layer is an affine mapping, its Lipschitz constant is the spectral
norm of the weight matrix, which can be controlled by multiplying
the weight by a constant. Therefore, if the contractive property is
found violated, i.e., $L\geq1$, we can correct it by replacing the
weights of each Conv layer in $R_{\Theta}$, i.e., $\Theta$, by $(\beta/L)^{\frac{1}{9}}\Theta$,
where $0\leq\beta<1$ is the desired Lipschitz constant of the NLE\footnote{The exponent $\frac{1}{9}$ is chosen because $R_{\Theta}(\cdot)$
has 9 Conv layers in total. Although there are residual connections
in the RBs, \cite{2021Heaton} has shown that the Lipschitz constant
can still be controlled in this way, as long as $\beta/L$ is close
to \nolinebreak 1, i.e., the contractive property is not seriously
violated. Actually, we observed that the correction was seldom triggered
during training. }. 

\addtolength{\topmargin}{0.02in}

\section{Simulation Results}

\begin{figure*}[tbh]
\begin{centering}
\subfloat[\label{fig:NMSE-SNR}]{\begin{centering}
\includegraphics[width=0.3\textwidth]{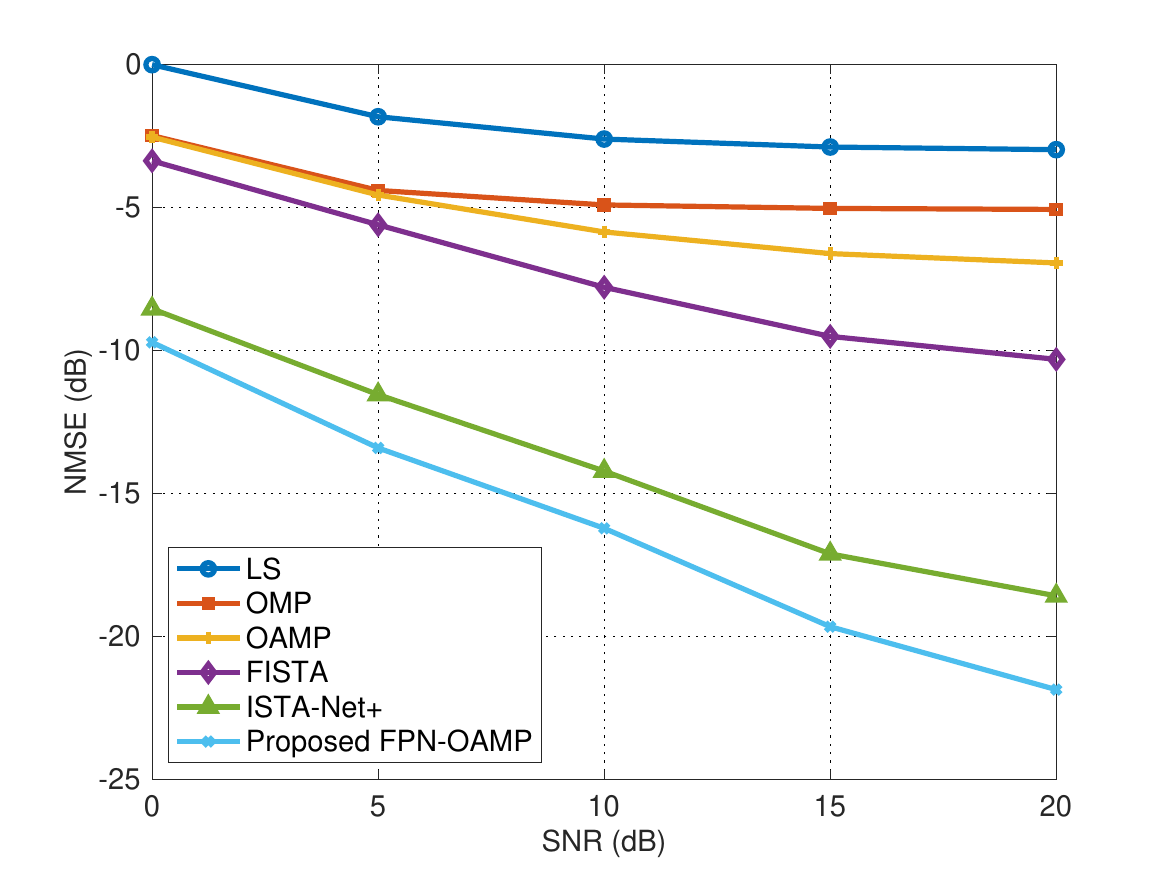}
\par\end{centering}
}\subfloat[\label{fig:NMSE-layer}]{\begin{centering}
\includegraphics[width=0.3\textwidth]{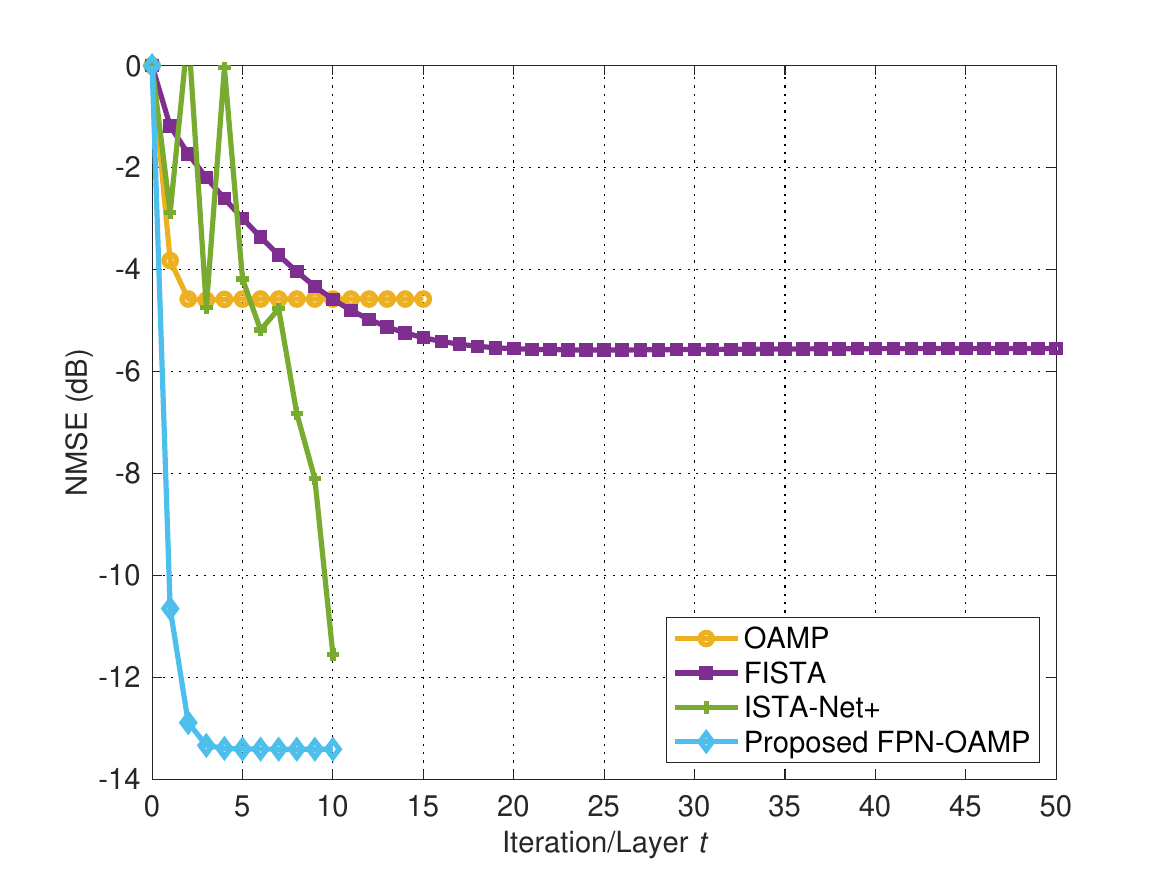}
\par\end{centering}
}\subfloat[\label{fig:Diffnorm-layer}]{\begin{centering}
\includegraphics[width=0.3\textwidth]{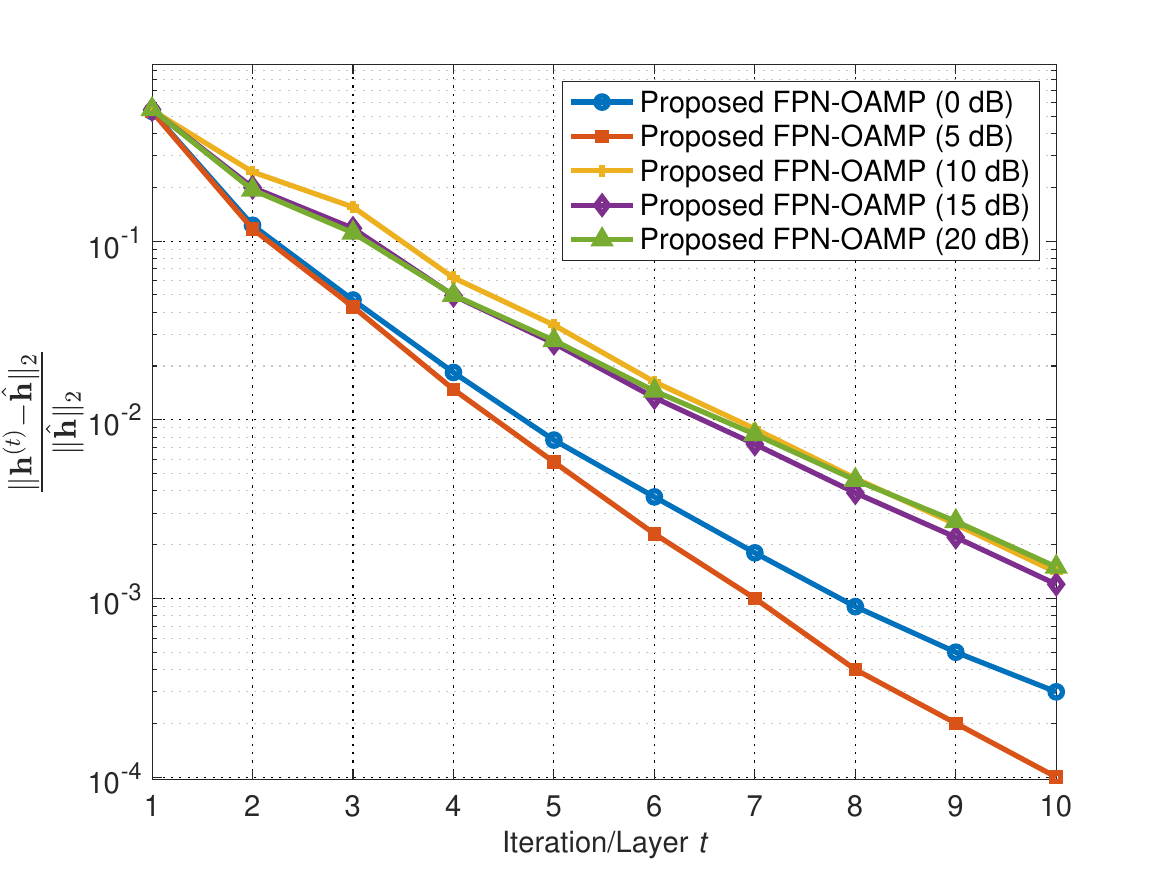}
\par\end{centering}
}\caption{Simulation results. (a) NMSE performance versus SNR. (b) NMSE evaluated
at iteration/layer $t$ when $\text{SNR}=5$ dB. (c) Normalized gap
between the intermediate estimate at iteration/layer $t$ and the
approximate fixed point of the proposed FPN-OAMP. \label{fig:Simulation-results}}
\par\end{centering}
\end{figure*}

This section provides simulation results to evaluate the performance
of the proposed FPN-OAMP method in a typical THz UM-MIMO system \cite{2021Dovelos}.
The key simulation parameters are summarized in Table \ref{tab:Key-Simulation-Parameters}.
Specifically, $r_{l}$ is set as a random variable spanning both far-
and near-field regions to model the hybrid-field propagation. The
performance metric is the NMSE, which is averaged over a testing dataset
with 5000 samples. Five benchmarks are adopted for comparison: 
\begin{itemize}
\item \textbf{LS}: Least squares. 
\item \textbf{OMP}: Orthogonal matching pursuit \cite{2016Lee,2021Dovelos,2022Wei}. 
\item \textbf{OAMP}: OAMP with the pseudoinverse LE \cite{2017Ma}. 
\item \textbf{FISTA}: Fast iterative soft thresholding algorithm \cite{2009Beck}. 
\item \textbf{ISTA-Net+}: State-of-the-art DU method based on the iterative
soft thresholding algorithm \cite{2018Zhang}. The number of layers
is fixed as 10, since a further increase is observed to offer only
negligible performance gain.
\end{itemize}
\begin{table}[t]
\caption{Key Simulation Parameters\label{tab:Key-Simulation-Parameters}}

\centering{}%
\begin{tabular}{|l|l|}
\hline 
\textbf{Parameter} & \textbf{Value}\tabularnewline
\hline 
Number of SAs / RF chains & $S=4$\tabularnewline
Number of AEs per SA & $\bar{S}=256$\tabularnewline
Number of BS antennas & $S\bar{S}=1024$\tabularnewline
Carrier frequency & $f_{c}=300$ GHz\tabularnewline
AE spacing & $d_{a}=5.0\times10^{-4}$ m\tabularnewline
SA spacing & $d_{\text{sub}}=5.6\times10^{-2}$ m\tabularnewline
Pilot length & $Q=128$\tabularnewline
Azimuth AoA & $\theta_{l}\sim\mathcal{U}(-\pi/2,\pi/2)$\tabularnewline
Elevation AoA & $\phi_{l}\sim\mathcal{U}(-\pi,\pi)$\tabularnewline
Angle of incidence & $\varphi_{\text{in},l}\sim\mathcal{U}(0,\pi/2)$\tabularnewline
Number of paths & $L=5$\tabularnewline
Rayleigh distance & $D_{\text{Rayleigh}}=20$ m\tabularnewline
LoS path length & $\text{\ensuremath{r_{1}=30}}$ m\tabularnewline
Scatterer distance ($l>1$) & $r_{l}\sim\mathcal{U}(10,25)$ m\tabularnewline
Time delay of LoS path & $\tau_{1}=100$ nsec\tabularnewline
Time delay of NLoS paths ($l>1$) & $\tau_{l}\sim\mathcal{U}(100,110)$ nsec\tabularnewline
Absorption coefficient & $k_{\text{abs}}=0.0033$ m$^{-1}$\tabularnewline
Refractive index & $n_{t}=2.24-j0.025$\tabularnewline
Roughness factor & $\sigma_{\text{rough}}=8.8\times10^{-5}$ m\tabularnewline
\hline 
\end{tabular}
\end{table}

We train two different sets of parameters for both ISTA-Net+ and the
proposed FPN-OAMP: one for the low SNR scenario (0 to 10 dB), and
the other for the high SNR scenario (10 to 20 dB). For each scenario,
we generate 80000, 5000, and 5000 samples for the purpose of training,
validation, and testing, respectively. The SNR of each sample is randomly
drawn based on its respective scenario. We train the networks for
150 epochs using the Adam optimizer with an initial learning rate
of 0.001 and a batch size of 128. The learning rate is reduced by
half after every 30 epochs. When training the FPN-OAMP, we set the
error tolerance $\epsilon$ as 0.01 and the maximum number of iteration
as 15 to accelerate the process. Note that in the testing stage, FPN-OAMP
can run for an arbitrary number of iterations, depending on the channel
condition and the error tolerance $\epsilon$. For fair comparison
with ISTA-Net+, during testing, we also set the the error tolerance
as 0.01, and the maximum number of iterations as 15. 

In Fig. \ref{fig:Simulation-results}(a), we present the NMSE performance
versus the average received SNR. It is observed that the proposed
FPN-OAMP significantly outperforms all five benchmarks under different
SNR levels. Compared with its base algorithm OAMP, the performance
gain of FPN-OAMP is as large as about 15 dB in terms of NMSE. This
indicates that the CNN component of FPN-OAMP can effectively identify
and exploit the complicated hybrid-field channel conditions. 

In Fig. \ref{fig:Simulation-results}(b), we illustrate the NMSE evaluated
at different iteration/layer $t$ when $\text{SNR}=5$ dB. LS and
OMP are not plotted since they do not produce intermediate results.
As observed, NMSE of the proposed FPN-OAMP and its base algorithm
OAMP converges rapidly within only 4 iterations, while FISTA converges
after about 20 iterations. Notably, the performance of FPN-OAMP at
iteration 2 has already outperformed the final performance of all
benchmark methods, which demonstrates its superior efficiency. By
contrast, the intermediate performance of the DU method, ISTA-Net+,
is very unstable and keeps fluctuating until layer 7. This is because
DU is a black-box solver which only seeks to optimize the final estimation
quality but does not explicitly control the internal dynamics like
the proposed FPN-OAMP. Additionally, the fixed number of layers limits
its ability to adapt to the changeable hybrid-field channel \nolinebreak conditions. 

Fig. \ref{fig:Simulation-results}(c) provides numerical verifications
for the linear convergence of FPN-OAMP under different SNR levels.
In logarithmic scale, we plot the normalized gap between the intermediate
estimate at iteration/layer $t$, i.e., $\mathbf{h}^{(t)}$, and the
approximate fixed point, i.e., $\mathbf{\hat{h}}$, defined by $\|\mathbf{h}^{(t)}-\mathbf{\hat{h}}\|_{2}/\|\mathbf{\hat{h}}\|_{2}$.
It is observed that the curves are all linear, which demonstrates
that the linear convergence property identified in Subsection \ref{subsec:Linear-Convergence}
is consistent under different SNR levels, and the training strategy
presented in Subsection \ref{subsec:Training-the-FPN-OAMP} is effective.
Although we plot the averaged results over the testing dataset for
illustration, it is worth noting that the linear convergence property
holds for each individual sample.

\section{Conclusions and Future Work}

In this paper, we proposed an efficient deep learning based hybrid-field
channel estimator for THz UM-MIMO. Significant performance gains are
then observed in comparison with state-of-the-art benchmarks. One
unique advantage is the linear convergence guarantee based on fixed
point theory, which is not available in the prevailing DU methods.
Besides, the computational complexity is adaptive, making it more
suitable for future wireless networks with complicated hybrid-field
channel conditions. Further extension of the proposed FPNs to other
important inverse problems in wireless communications, such as data
detection, is a promising future direction. 

\bibliographystyle{IEEEtran}
\bibliography{references}

\end{document}